# Orbital anomalous Hall effect in the few-layer Weyl semimetal TaIrTe$_4$

An-Qi Wang,[1,*] Dong Li,[1,2,*] Tong-Yang Zhao,[1,*] Xing-Yu Liu,[1] Jiantian Zhang,[3] Xin Liao,[1] Qing Yin,[1]
Zhen-Cun Pan,[1] Peng Yu,[3] and Zhi-Min Liao[1,4,†]

[1]*State Key Laboratory for Mesoscopic Physics and Frontiers Science Center for Nano-optoelectronics,
School of Physics, Peking University, Beijing 100871, China*
[2]*Academy for Advanced Interdisciplinary Studies, Peking University, Beijing 100871, China*
[3]*State Key Laboratory of Optoelectronic Materials and Technologies, School of Materials Science and Engineering,
Sun Yat-sen University, Guangzhou 510275, China*
[4]*Hefei National Laboratory, Hefei 230088, China*

We report on the observation of the linear anomalous Hall effect (AHE) in the nonmagnetic Weyl semimetal TaIrTe$_4$. This is achieved by applying a direct current $I^{dc}$ and an alternating current $I^\omega$ ($I^\omega \ll I^{dc}$) in TaIrTe$_4$, where the former induces time-reversal symmetry breaking and the latter probes the triggered AHE. The anomalous Hall resistance $V_H^\omega/I^\omega$ shows a linear dependence on $I^{dc}$ and changes sign with the polarity of $I^{dc}$. In temperature-dependent measurements, $V_H^\omega/I^\omega$ also experiences a sign reversal at ∼100 K, consistent with the temperature-dependent nonlinear Hall effect (NLHE). Furthermore, in measurements involving only dc transport, the dc Hall voltage exhibits a quadratic relationship with $I^{dc}$. When the $I^{dc}$ direction is reversed, the Hall resistance changes sign, demonstrating a colossal nonreciprocal Hall effect (NRHE). Our theoretical calculations suggest that the observed linear AHE, NLHE, and NRHE all dominantly originate from the current-induced orbital magnetization compared to the minor spin contribution. This work provides deep insights into the orbital magnetoelectric effect and nonlinear Hall response, promising precise electric control of out-of-plane polarized orbit flow.

## I. INTRODUCTION

In the past decade, there has been a surge in research focused on exploring the nonlinear Hall effect (NLHE) as an efficient tool for uncovering inherent properties of electronic states like the quantum geometric tensor [1–8], including Berry curvature [9–13] and quantum metrics [14–17]. Theoretical analysis of nonlinear conductivity tensor suggests that second-order Hall response can emerge in materials with time-reversal symmetry [12,13,18]. Experimental progress on second-order NLHE has been achieved in systems including two-dimensional transition metal dichalcogenides [19–26], Weyl semimetal TaIrTe$_4$ [27], corrugated [28] or twisted [29] bilayer graphene, and topological materials [16,17,30–32]. Moreover, third-order Hall response induced by a mechanism known as Berry connection polarizability [33,34] has also been demonstrated in topological semimetals such as MoTe$_2$ [35], WTe$_2$ [35,36], TaIrTe$_4$ [37], and Cd$_3$As$_2$ [38]. In contrast, the first-order (linear) anomalous Hall effect (AHE) rigorously requires the breaking of time-reversal symmetry according to the Onsager reciprocity relation. It describes the production of transverse voltage upon applying a longitudinal current in the absence of an external magnetic field [11,39–41]. In the general cases, the AHE typically occurs in systems with spontaneous magnetization [42–47]. In these materials, there are unequal numbers of spin-up and spin-down electrons, which are deflected towards opposite sides due to intrinsic Berry curvature and extrinsic spin-dependent scatterings, resulting in a nonzero transverse voltage [Fig. 1(a)]. This scenario can be extended to the orbital analog, where orbital magnetic moment dependent deflections contribute to the AHE [18,48,49], i.e., orbital AHE [Fig. 1(b)]. Despite its common presence together with AHE, the spontaneous magnetism is not a necessity for AHE to be observed [50–53]. In nonmagnetic materials, it is possible to induce the polarization of electron spin or orbital moments by a charge current with the time-reversal symmetry breaking, where the linear AHE is anticipated.

Here we report the observation of current-induced linear AHE based on two-dimensional nonmagnetic materials. In two-dimensional crystals, the orbital magnetic moment of electrons is bound to align out-of-plane due to the dimension constraint [9,54–56], naturally enforcing the orbital current to flow in-plane orthogonally to bias current [57–61], as indicated in Fig. 1(b). Therefore, if there is a polarization of the orbital magnetic moment in the bias current, an observable Hall signal will be generated. The presence of Berry curvature dipole $\boldsymbol{D}$, the dipole moment of Berry curvature over the occupied states, would trigger an out-of-plane polarization of orbital magnetic moments $M_{orb}^z$ [22,23] and break the time-reversal symmetry under a charge current $I$ [Fig. 1(c)], formulated as $M_{orb}^z \propto (DI\cos\theta)\hat{z}$ with $\theta$ being the angle between the charge current and the Berry curvature dipole. The current-induced out-of-plane orbital magnetization causes an imbalance in the transverse motion of Bloch electrons, generating nonzero Hall signals under a longitudinal electric field excitation. If merely a single current $I$ is applied, both the polarization and detection depend linearly on $I$, leading to the second-order NLHE as reported in previous literatures [19–29]. Here, we apply a moderate fixed dc current to serve as the polarization generator and a small ac current to trigger the Hall signal, respectively, where the measured ac Hall voltage exhibits a linear dependence on the ac current resembling a regular first-order AHE. The dc+ac experimental method allows us to independently control the generation and detection of orbital magnetization, helping to unveil the correlation between nonlinear charge response and orbital magnetization (see Appendix A).

---
[*]These authors contributed equally to this work.
[†]Contact author: liaozm@pku.edu.cn

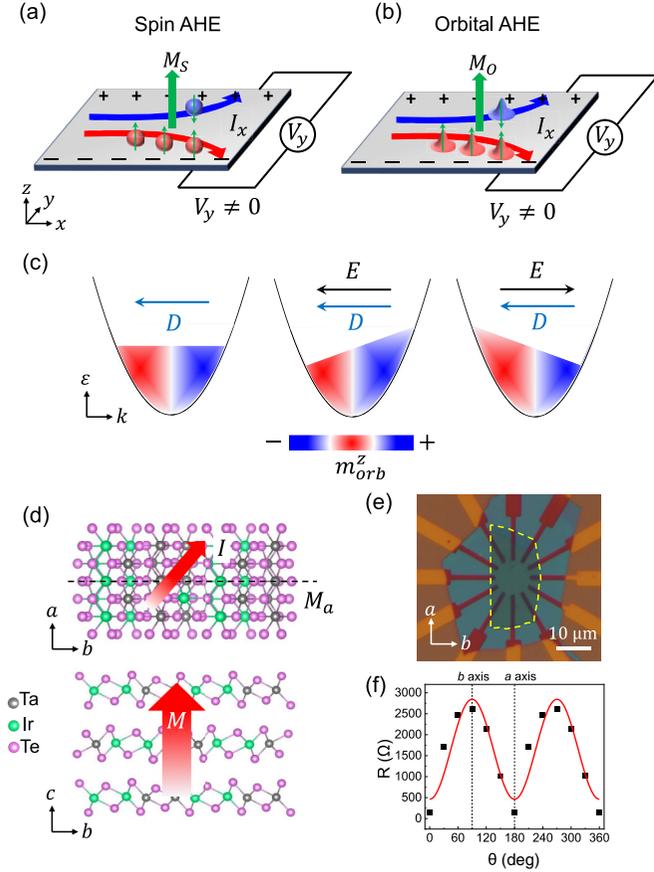

FIG. 1. Crystal structure and basic characterization of few-layer TaIrTe$_4$. (a) Schematic of the anomalous Hall effect in a magnetic metal. $M_S$ is the spin magnetization; $I_x$ and $V_y$ are the bias current and Hall voltage, respectively. Electrons with opposite spin magnetic moments are denoted by red and blue, respectively. The green arrows on the electrons represent the direction of the spin magnetic moments. (b) Schematic of the anomalous Hall effect due to orbital magnetization, denoted as $M_O$, i.e., the orbital anomalous Hall effect. Bloch electrons with opposite orbital magnetic moments are denoted by red and blue, respectively. The green arrows on the electron wave packets represent the direction of orbital magnetic moments. (c) Mechanism of current-induced orbital magnetization. The presence of Berry curvature dipole $D$ leads to a unique orbital texture, where the electron momentum $k$ is locked to the orbital magnetic moment $m_{orb}^z$. In the absence of an electric field, the number of electrons with upward $m_{orb}^z$ is equal to that with downward $m_{orb}^z$, resulting in no polarization of the orbital magnetic moment. When applying an electric field $E$, this balance is broken, leading to a net polarization of $m_{orb}^z$, i.e., orbital magnetization. (d) Top (upper panel) and side (lower panel) views of the crystal structure of few-layer TaIrTe$_4$. When an in-plane current $I$ is applied, an out-of-plane orbital magnetization $M$ is generated, with $D$ along the $a$ axis. (e) Optical image of the few-layer TaIrTe$_4$ device. The TaIrTe$_4$ flake is outlined by a yellow dashed box. (f) Resistance anisotropy in the TaIrTe$_4$ device. The $\theta$ is defined as the angle relative to the $a$ axis; for instance, $R(\theta = 90°)$ corresponds to the resistance along the $b$ axis. The resistances were measured via the four-probe method.

## II. EXPERIMENTAL METHOD

We have chosen few-layer $T_d$-TaIrTe$_4$ for the investigation of current-induced orbital magnetization and linear AHE [Fig. 1(d)]. In contrast to its bulk counterpart, the thin TaIrTe$_4$ layer exhibits mirror symmetry $M_a$ but lacks glide mirror symmetry $\tilde{M}_b$ due to disrupted translation symmetry along the $c$ axis [27,37,62]. This material showcases a prominent Berry curvature dipole $D$ perpendicular to the mirror plane (along the crystal $a$ axis), which has been revealed by the observation of robust NLHE [27]. When a bias current ($I$) is applied in the $a$-$b$ plane, it generates an out-of-plane orbital magnetization ($M$) along the $c$ axis [Fig. 1(d)] with breaking of the time-reversal symmetry. On the other hand, despite the presence of spin-orbit coupling, the current-induced spin magnetization is vanishingly small compared to the orbital magnetization in few-layer TaIrTe$_4$ (see theoretical calculation in Fig. 2). A typical circular disk device based on an 8 nm-thick TaIrTe$_4$ flake is shown in Fig. 1(e). The crystalline orientations of the TaIrTe$_4$ device are identified by its long and straight edges and further confirmed using angle-resolved polarized Raman spectroscopy (see Fig. 5 and Appendix B). The resistance in various directions exhibits a twofold angular dependence [Fig. 1(f)], in accordance with the crystal symmetry of few-layer TaIrTe$_4$ [63]. The angle-dependent resistance can be fitted by the formula $R(\theta) = R_a\cos^2\theta + R_b\sin^2\theta$, where $R_a$ and $R_b$ represent the resistance along the $a$ and $b$ axes, respectively. The fitting yields a resistance anisotropy $r = R_a/R_b \approx 0.15$, similar to previous results on TaIrTe$_4$ [27,37,64].

## III. RESULTS AND DISCUSSION

Before electronic transport measurements, we performed theoretical calculations of the orbital (spin) textures and magnetoelectric susceptibility in TaIrTe$_4$ (see Appendix C for more details), which remain elusive in previous works [27,37,64–66]. Figure 2(a) shows the calculated energy band structure of few-layer TaIrTe$_4$. Both the electron and hole would contribute to electronic transport (two-carrier transport) near the charge neutrality point [indicated by the red dashed line in Fig. 2(a)], consistent with the previous theoretical and experimental results [67–70]. The spatial distribution of orbital magnetic moment $m_{orb}^z$ in the momentum space is calculated as shown in Fig. 2(b). Due to the conserved time-reversal symmetry, $m_{orb}^z(k) = -m_{orb}^z(-k)$, and the total orbital moment is zero, exhibiting no magnetization. Recent studies suggest that spin-orbit coupling and low symmetry allow $z$ spins in TaIrTe$_4$ [64,71]. To compare with the orbital moment, we also performed calculations to obtain the spin moment texture of TaIrTe$_4$, as shown in Fig. 2(c). The amplitude of the orbital moment reaches up to 60 $\mu_b$ near the $k$ point $(\pm 0.1, 0)$ Å$^{-1}$, which is much larger than that of the spin moment. To characterize the efficiency of current-induced out-of-plane magnetization, we introduce the concept of magnetoelectric susceptibility $\alpha_{ij}$ (see Appendix C). Figure 2(d) compares the orbital and spin magnetoelectric susceptibility, denoted by $\alpha_{xz}^{\mathrm{orb}}$ and $\alpha_{xz}^s$, as a function of Fermi level $\mu$. The Fermi level of the TaIrTe$_4$ flake is estimated to be near $\mu = 0.037$ eV [indicated by the blue dashed line in Fig. 2(d)] according to the electron and hole carrier densities obtained from the Hall measurement. At this Fermi level, the induced orbital magnetization $M_{\mathrm{orb}}^z$ is found to be much stronger than the induced spin magnetization $M_s^z$ in the presence of a bias current. In the scenario of AHE, the anomalous Hall resistance is generally believed to scale linearly with the magnetization $M^z$, formulated as $R_H = \gamma M^z$, where $\gamma$ is a sample-dependent quantity linked with intrinsic Berry curvature and extrinsic disorder scatterings [41]. Considering the relation $M_{\mathrm{orb}}^z \gg M_s^z$, the current-induced orbital magnetization is anticipated to make a dominant contribution (over spin) to the anomalous Hall signals ($R_H$) in the studied TaIrTe$_4$ flake device.

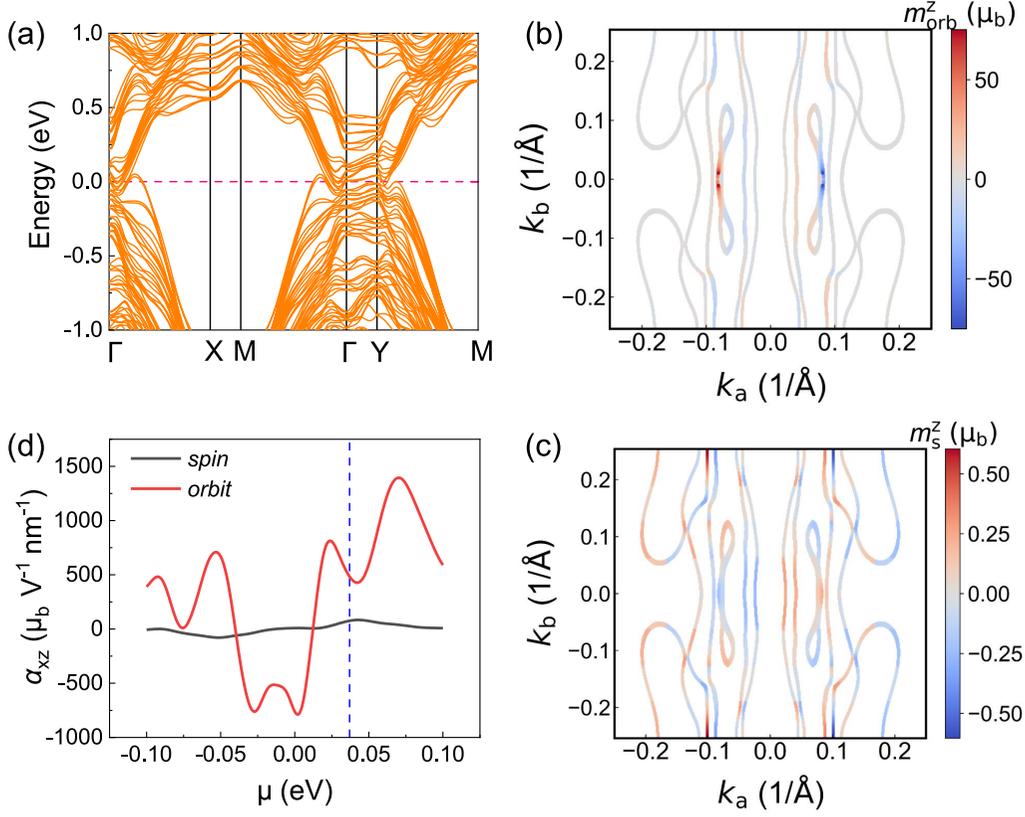

FIG. 2. Theoretical calculations of out-of-plane orbital and spin magnetoelectric effect in TaIrTe$_4$. (a) Energy band structure of pentalayer $T_d$-TaIrTe$_4$. (b), (c) Calculated momentum $k$-resolved distribution of magnetic moment for (b) orbit and (c) spin at $\mu = 0$, as indicated by the red dashed line in (a). $m^z$ denotes the out-of-plane component of magnetic moment. $\mu_b$ is the Bohr magneton. (d) The theoretical magnetoelectric susceptibility $\alpha_{xz}$ as a function of chemical potential $\mu$. The $\alpha_{xz}^{\mathrm{orb}}$ ($\alpha_{xz}^s$) reflects the strength of induced out-of-plane orbital (spin) magnetization per unit electric field $E^x$. The temperature used for the calculations is 50 K. The coordinates $x, y, z$ correspond to the $a, b,$ and $c$ axes of TaIrTe$_4$, respectively.

Experimentally, we apply a dc current ($I^{\mathrm{dc}}$) in the $a$-$b$ plane to induce an out-of-plane magnetization. A probing ac current ($I^\omega$) at a frequency $\omega$ is applied along the $a$ axis, and the transverse Hall voltage ($V_H^\omega$) is detected along the $b$ axis [Fig. 3(a)]. The condition $I^\omega \ll I^{\mathrm{dc}}$ is met to primarily determine the magnetization by $I^{\mathrm{dc}}$. With a fixed $I^{\mathrm{dc}}$ applied along the $a$ axis, $V_H^\omega$ displays a linear relationship with $I^\omega$, as shown in Fig. 3(b). Furthermore, the slope of the $V_H^\omega - I^\omega$ curve increases as $I^{\mathrm{dc}}$ is augmented. In Fig. 3(c), the Hall resistance ($R_H^\omega = V_H^\omega/I^\omega$) is plotted against $I^{\mathrm{dc}}$, revealing a linear proportionality. This is expected since the Hall resistance is proportional to the induced out-of-plane magnetization which is mainly contributed by orbital magnetization and linear to the $I^{\mathrm{dc}}$ in the presence of a nonzero $D$. In contrast to the case of $I^{\mathrm{dc}}$ along the $a$ axis, the $R_H^\omega$ is found to be vanishingly small when the bias current $I^{\mathrm{dc}}$ is along the $b$ axis (see Fig. 7 and Appendix D). This can be well understood by considering that the induced orbital magnetization is negligible with $I^{\mathrm{dc}}$ along the $b$ axis where the current direction is perpendicular to $D$.

Next, we investigate the temperature dependence of AHE in TaIrTe$_4$ with $I^{\mathrm{dc}}$ along the $a$ axis [Fig. 4(a)]. As the temperature rises, $V_H^\omega$ gradually diminishes and even changes signs. In Fig. 4(b), we plot $R_H^\omega$ as a function of temperature. $R_H^\omega$ decreases with increasing temperature until 100 K, accompanied by a sign reversal, and then gradually approaches zero at 250 K. In our device, the NLHE also exhibits similar temperature dependence (see Appendix E), with a change in sign observed above 100 K [Fig. 4(c)]. Prior reports have shown that increasing temperatures induces a shift in the Fermi level of TaIrTe$_4$, leading to a change in both the direction of the Berry curvature dipole and the sign of second-order NLHE [27]. Both the flipping of the direction of $D$ and extrinsic disorder scatterings [27] could lead to the observed variation in the sign of $R_H^{\omega}$.

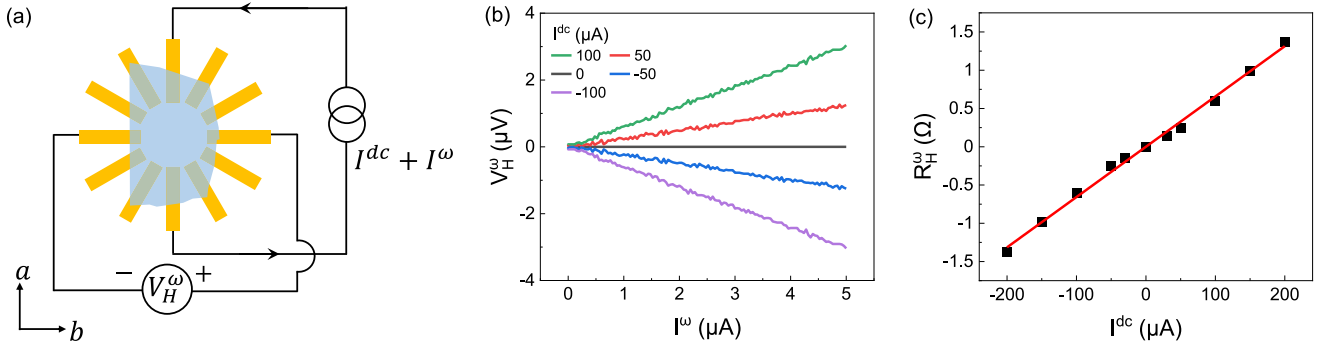

FIG. 3. The linear AHE observed in the TaIrTe$_4$ device. (a) Illustration of measurement configuration for the current-induced AHE. A combination of dc current $I^{dc}$ and ac current $I^{\omega}$ is injected along the $a$ axis while recording the transverse Hall voltage $V_H^{\omega}$. The applied $I^{dc}$ is to induce out-of-plane magnetization, and the $I^{\omega}$ ($I^{\omega} \ll I^{dc}$) serves as a probing current to measure the AHE. (b) The measured $V_H^{\omega}$ as a function of $I^{\omega}$ under different $I^{dc}$. (c) The Hall resistance $R_H^{\omega}$ $\left(\frac{V_H^{\omega}}{I^{\omega}}\right)$ vs $I^{dc}$.

The AHE can also be demonstrated in pure dc measurements [Fig. 4(d)]. In this regime, the transverse Hall voltage will exhibit a quadratic dependence on the applied dc bias current, as dictated by the mechanism responsible for current-induced out-of-plane magnetization and AHE. Our experimental setup involves injecting a dc current ($I^{dc}$) into the TaIrTe$_4$ flake along the $a$ axis while measuring the transverse Hall voltage ($V_H^{dc}$) across the electrodes along the $b$ axis [see the inset in Fig. 4(f)]. By symmetrizing the measured data (see Appendixes A and F), the portion of the quadratic dependence of $V_H^{dc}$ on the current $I^{dc}$, denoted as $V_H^s$, is obtained. The results obtained at various temperatures are summarized in Fig. 4(d). The quadratic $V_H^s - I^{dc}$ relation [Fig. 4(e)] means that the transverse resistance changes sign upon switching the current direction, manifesting a pronounced nonreciprocal Hall effect (NRHE) (see Appendix F). Furthermore, this nonreciprocal Hall signal reverses sign as the temperature rises above 100 K [Fig. 4(f)], consistent with the observations in Figs. 4(b) and 4(c). The linear AHE, second-order NLHE, and NRHE in our TaIrTe$_4$ device should stem from a common origin: current-induced orbital magnetization via a nonzero Berry curvature dipole, and the related experimental results are quantitatively consistent with each other (see Fig. 11 and Appendix H).

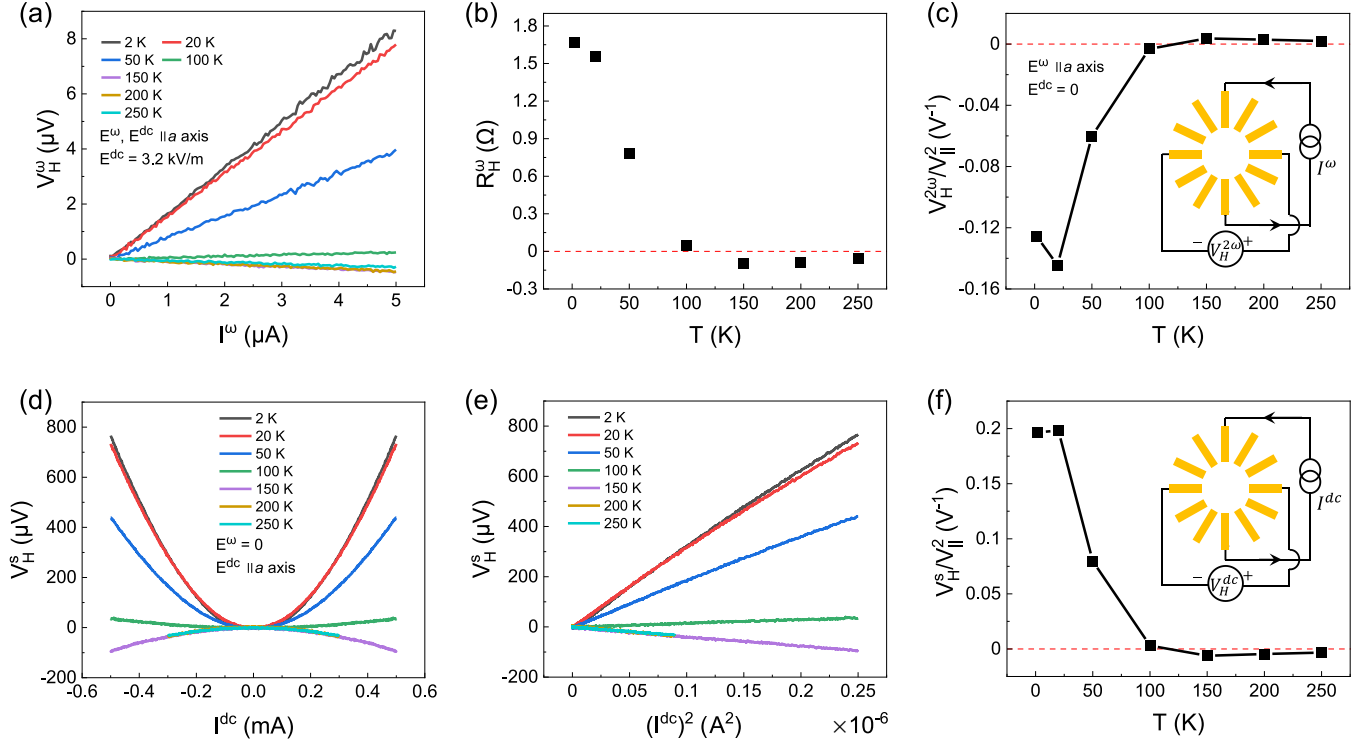

FIG. 4. Temperature dependence of linear AHE, second-order NLHE, and NRHE in TaIrTe$_4$. (a) $V_H^{\omega} - I^{\omega}$ curves measured under various temperatures. The dc current $I^{dc}$ is directed along the $a$ axis with a constant electric field $E^{dc}$ of 3.2 kV/m for all temperatures. (b) The Hall resistance $R_H^{\omega}$ as a function of temperature, corresponding to the data in (a). A sign reversal of $R_H^{\omega}$ is evident above 100 K. (c) Temperature dependence of the NLHE measured using the configuration illustrated in the inset. The nonlinear Hall signal $V_H^{2\omega}$ is scaled by the factor $V_\parallel^2$ to normalize the ability to generate NLHE under various temperatures, where $V_\parallel$ is the longitudinal voltage. The dc current $I^{dc}$ is set to zero for NLHE measurement, while the ac current $I^{\omega}$ remains injecting along the $a$ axis. (d)–(f) NRHE measured via a pure dc method, where ac current $I^{\omega}$ is zero and dc current $I^{dc}$ is applied along the $a$ axis, and the transverse Hall voltage $V_H^{dc}$ is recorded (as depicted in the right panel). (d) The symmetric component of $V_H^{dc}$, denoted as $V_H^s$, as a function of bias current $I^{dc}$. $V_H^s$ is obtained by symmetrizing the original $V_H^{dc} - I^{dc}$ data. (e) Linear scaling of $V_H^s$ with the square of $I^{dc}$ under varying temperatures. (f) The slope $V_H^s/V_\parallel^2$ vs the temperature $T$, where $V_\parallel$ is the longitudinal voltage.

## IV. CONCLUSIONS

We have theoretically and experimentally investigated the out-of-plane magnetoelectric effect in few-layer Weyl semimetal TaIrTe$_4$. Theoretical calculations indicate that the current-induced orbital magnetization is dominant over the spin magnetization in TaIrTe$_4$. Experimentally, we apply dc current to induce an out-of-plane magnetization, and employ the ac current to probe the induced AHE. The anomalous Hall resistance $R_H^\omega$ ($V_H^\omega/I^\omega$) is found to scale linearly with the $I^{dc}$ and change sign with the polarity of $I^{dc}$ as expected from the proportional relation between orbital magnetization and bias current. Moreover, we observe a sign reversal of $R_H^\omega$ as temperature is varied, consistent with the temperature-dependent sign change of the orbital magnetization. We also demonstrate a nonlinear Hall effect in pure dc measurements, which more clearly showcases the directional transverse motion of carriers with out-of-plane orbital polarization compared to second-harmonic ac measurements.

## ACKNOWLEDGMENTS

This work was supported by the National Natural Science Foundation of China (Grants No. 62425401, No. 62321004, and No. 12204016), and the Innovation Program for Quantum Science and Technology (Grant No. 2021ZD0302403). P.Y. was supported by the National Natural Science Foundation of China (Grant No. 22175203) and the Natural Science Foundation of Guangdong Province (Grant No. 2022B1515020065).

## APPENDIX A: DEVICE FABRICATION, MEASUREMENT METHODS, AND DATA PROCESSING

### 1. Sample growth and device fabrication

Few-layer TaIrTe$_4$ (8 nm thick) and hBN ($\sim 10-20$ nm thick) flakes were obtained via the mechanical exfoliation method from bulk crystals (HQ Graphene). The process involved exfoliating the TaIrTe$_4$ crystal with Scotch Tape and transferring it onto a polydimethylsiloxane (PDMS) substrate. Subsequently, it was transferred onto a SiO$_2$/Si substrate that had been precleaned. Circular disk-shaped Ti/Au electrodes (2 nm/8 nm thick) were patterned via $e$-beam lithography and deposited via an $e$-beam evaporation process onto a SiO$_2$/Si substrate. The exfoliated hBN and TaIrTe$_4$ flakes were successively picked up using the dry transfer method and then placed onto the prepatterned Ti/Au electrodes. The long and straight edge (indication of the crystal $a$ axis) of the TaIrTe$_4$ was aligned with one pair of the Ti/Au electrodes. To avoid possible degradation, the exfoliation and transfer processes were both performed in a nitrogen-filled glove box.

### 2. Transport measurements

Transport measurements were performed in a commercial Oxford cryostat system with a base temperature $\sim 1.4$ K. To study the current-induced out-of-plane magnetization and AHE, two measurement methods were employed. In the first method, where both dc and ac currents were simultaneously introduced into the sample (Fig. 3), a dc current ($I^{dc}$) was applied to TaIrTe$_4$ in the $ab$ plane using a Keithley 2400 Source Meter. Meanwhile, an ac probe current $I^\omega$ ($I^\omega \ll I^{dc}$) with frequency $\omega$ was applied along the $a$ axis (SR830), and the first-harmonic transverse Hall voltage $V_H^\omega$ was measured via the lock-in technique. For the second method, which involved pure dc measurements [Fig. 4(f)], a dc current ($I^{dc}$) was applied along the $a$ axis, while the transverse Hall voltage ($V_H^{dc}$) was captured across the electrodes along the $b$ axis. When studying the NLHE of TaIrTe$_4$, only the ac current was applied, with $I^\omega$ injected into the sample and the second-harmonic transverse voltage ($V_H^{2\omega}$) recorded. The driving frequency of the ac current was set at 17.777 Hz unless otherwise specified.

The dc+ac approach exhibits superiority over the pure ac measurement in unveiling the underlying mechanism of the out-of-plane magnetoelectric effect. For the pure ac measurement, the electric current for generating out-of-plane magnetization and detecting the triggered Hall signals always stays the same. By contrast, the dc+ac approach allows one to independently control the two currents, which is helpful to study the generation and the detection process in greater depth. For example, we can apply dc and ac current in different directions to investigate the angle dependence of current-induced magnetization (Fig. 7), which is obviously not accessible in the pure ac measurement. The dc+ac method provides an improved measurement technique allowing elaborate characterization of the current-induced out-of-plane magnetization and nonlinear charge response.

### 3. Antisymmetrization and symmetrization of the measured data

In the measurement configuration combining dc and ac current sources, the dc current is used to produce an out-of-plane magnetization, while the ac current serves as the probe current to measure the corresponding anomalous Hall effect. The sign of the Hall voltage $V_H^\omega$ should follow the sign of the dc current induced magnetization $M$. Consequently, when the direction of dc current $I^{dc}$ is reversed, the sign of the corresponding Hall voltage $V_H^\omega$ also reverses due to the inversion of $M$. During the data analysis, $V_H^\omega$ for a given $I^{dc}$ was initially subjected to antisymmetrization between opposing current directions, expressed as $[V_H^\omega(I^{dc}) - V_H^\omega(-I^{dc})]/2$. This antisymmetrization step is essential to eliminating the influence of longitudinal-transverse coupling and the thermoelectric effect induced by Joule heating. Each of these factors introduces a symmetric term in $V_H^\omega$ concerning $I^{dc}$.

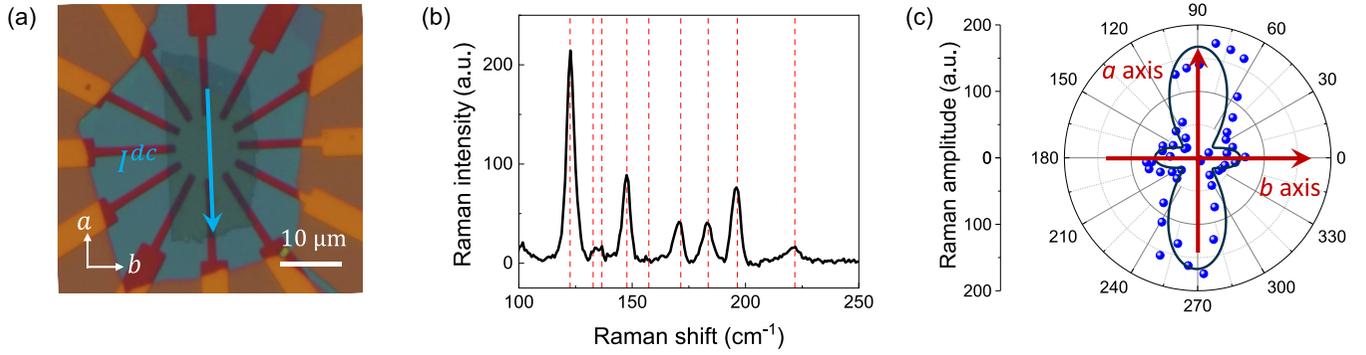

FIG. 5. Polarized Raman spectroscopy of TaIrTe$_4$ to determine the crystalline orientation. (a) Optical image of the studied TaIrTe$_4$ device. The crystalline $a$ and $b$ axes, denoted by white arrows, are determined by polarized Raman spectroscopy. The applied direct current $I^{dc}$ (light blue arrow) is approximately along the $a$ axis to induce the orbital magnetization. (b) Unpolarized Raman spectrum measured at the TaIrTe$_4$ flake. The position of nine Raman peaks is indicated by the red dashed lines. (c) Polar plot of the Raman spectrum (showing modes at $\sim$148 cm$^{-1}$) as a function of polarization direction. The polarization angle 90° corresponds to the $a$ axis, where the maximum intensity is detected.

In the pure dc method for detecting the AHE, only a dc current $I^{dc}$ was applied to the sample along the $a$ axis, and the corresponding transverse voltage $V_H^{dc}$ along the $b$ axis was measured. However, in practical experiments, the measured $V_H^{dc}$ often comprises both an antisymmetric component and a symmetric component. To reveal the AHE mechanism in which the Hall voltage exhibits a quadratic dependence on $I^{dc}$, the symmetric component of $V_H^{dc}$, denoted as $V_H^s$, was extracted through $[V_H^{dc}(I^{dc}) + V_H^{dc}(-I^{dc})]/2$.

## APPENDIX B: RAMAN SPECTROSCOPY MEASUREMENTS TO DETERMINE THE CRYSTALLINE ORIENTATION

Angle-resolved polarized Raman spectroscopy provides a useful tool to determine the crystalline orientation of the TaIrTe$_4$ flake. Figure 5(a) shows the optical image of the TaIrTe$_4$ device studied in the main text. Typical unpolarized Raman spectroscopy is shown in Fig. 5(b), where nine peaks are captured, consistent with the previous report of few-layer $T_d$-TaIrTe$_4$ flake [63]. We collected polarized Raman spectra in a parallel polarization configuration. Previous reports suggest that the Raman mode at 148 cm$^{-1}$ achieves a maximum amplitude when the excitation laser polarization is along the crystal $a$ axis [63]. Figure 5(c) exhibits the intensity of the Raman mode (148 cm$^{-1}$) as a function of polarization angle. It can be seen that the polarization angle 90° (0°) corresponds to the $a$ ($b$) axis of the TaIrTe$_4$ flake. The crystal $a$ and $b$ axes are indicated by white arrows in Fig. 5(a), where the bias current $I^{dc}$ (light blue arrow) is approximately parallel to the $a$ axis.

## APPENDIX C: THEORETICAL CALCULATION OF ORBITAL AND SPIN MAGNETIZATION

First-principle calculations were performed to reveal the orbital (spin) texture and corresponding magnetoelectric susceptibility of TaIrTe$_4$. We constructed a density functional theory (DFT) based tight-binding model Hamilton of the $T_d$-TaIrTe$_4$ slab, where the tight-binding model matrix elements are calculated by projecting onto the Wannier orbitals [72,73]. The $d$ orbitals of the Ta atoms, the $d$ orbitals of the Ir atoms, and the $p$ orbitals of the Te atoms were used to construct Wannier functions, without performing the procedure for maximizing localization.

### 1. Band structure and carrier density

The band structure of pentalayer $T_d$-TaIrTe$_4$ calculated from the DFT calculations is shown in Fig. 2(a) in the main text. The corresponding carrier densities of electrons ($n_e$) and holes ($n_h$) are given in Fig. 6(a). The carrier densities $n_e$ and $n_h$ are calculated by [27]

$$n_e = \int_{\epsilon_c}^{\infty} g_e(E) f_0(E - \mu_F) dE,$$
$$n_h = \int_{-\infty}^{\epsilon_v} g_h(E) f_0(\mu_F - E) dE, \quad \text{(C1)}$$

where $\epsilon_c$ and $\epsilon_v$ are the energies of the conduction band minimum and valence band maximum, respectively; $\mu_F$ is the Fermi energy; $g_e$ ($g_h$) is the density of states of electrons (holes); and $f_0$ is the Fermi-Dirac distribution.

### 2. Calculations of orbital and spin textures

We can calculate the orbital and spin magnetic moment of each Bloch state according to the following formulas [9,74]:

$$\vec{m}_{\text{orb}}(\vec{k}) = \frac{e}{2\hbar} \text{Im} \langle \partial_{\vec{k}} u(\vec{k}) | \times [H(\vec{k}) - \varepsilon(\vec{k})] | \partial_{\vec{k}} u(\vec{k}) \rangle, \quad \text{(C2)}$$

$$\vec{m}_s(\vec{k}) = -\langle \partial_{\vec{k}} u(\vec{k}) | \tfrac{1}{2} g \mu_b \sigma | \partial_{\vec{k}} u(\vec{k}) \rangle, \quad \text{(C3)}$$

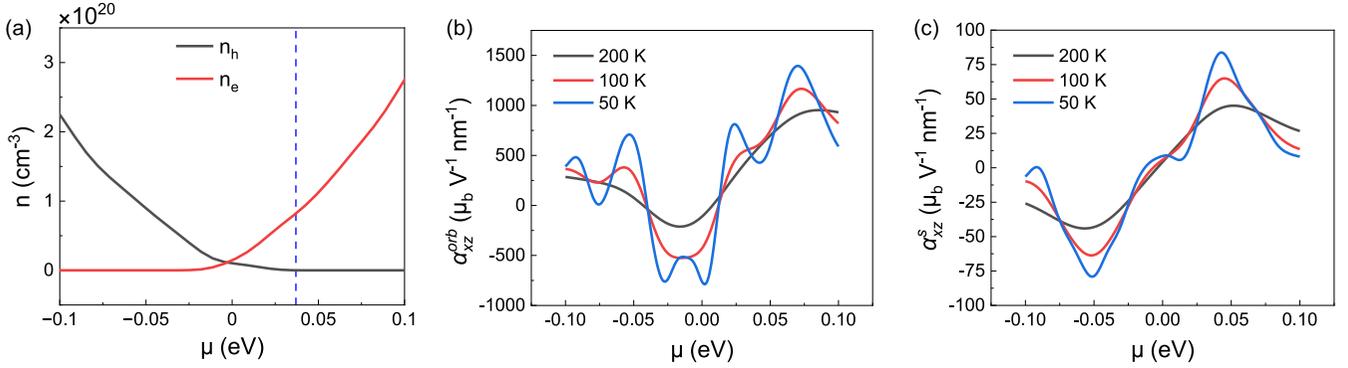

FIG. 6. Calculations of carrier density and magnetoelectric susceptibility. (a) The corresponding carrier densities of electrons ($n_e$) and holes ($n_h$) in the pentalayer $T_d$-TaIrTe$_4$. The blue dashed line denotes the position of the Fermi level in our TaIrTe$_4$ device ($\mu = 0.037$ eV). For the calculations of $n_e$ and $n_h$, the temperature is set at 0 K. (b), (c) The magnetoelectric susceptibility for the orbital moment ($\alpha_{xz}^{\text{orb}}$) and spin moment ($\alpha_{xz}^s$) at different temperatures, respectively.

where $e$ is the electron charge, $\hbar$ is the reduced Planck's constant, $u(\vec{k})$ is the periodic part of the Bloch state, $g$ is the Landé $g$ factor, and $\mu_b$ is the Bohr magneton. Utilizing Eqs. (C2) and (C3), we can obtain the distribution of local orbital and spin moments in the momentum space at a given Fermi level. Figure 2 provides the color plot of orbital and spin textures calculated at $\mu = 0$.

### 3. Calculations of magnetoelectric susceptibility $\alpha_{ij}$

The electric field induced magnetization can be described as $M_j = \sum_i \alpha_{ij} E_i$, where $i, j = x, y, z$; $E_i$ is the external electric field; and $\alpha_{ij}$ is the magnetoelectric susceptibility [75]. To calculate $\alpha_{ij}$ for TaIrTe$_4$, we can use the linear response theory [76,77], which gives

$$\alpha_{ij} = -\tau \frac{e}{\hbar} \frac{1}{(2\pi)^2} \int_{\text{BZ}} d^2k\, v^i(\vec{k}) m^j(\vec{k}) \left( \frac{\partial f_0}{\partial \varepsilon} \right)_{\varepsilon = \varepsilon(\vec{k})}, \quad \text{(C4)}$$

where $f_0$ is the Fermi-Dirac distribution function, $v(\vec{k})$ is the group velocity, $e$ is the electron charge, $\tau$ is the effective scattering time, $\varepsilon(\vec{k})$ is the band energy of the Bloch state, and $m(\vec{k})$ is the total magnetic moment. For the spin magnetic moment, $m_s(\vec{k}) = -\langle \partial_{\vec{k}} u(\vec{k}) | \frac{1}{2} g \mu_b \sigma | \partial_{\vec{k}} u(\vec{k}) \rangle$, as shown by Eq. (C3), while for the orbital magnetic moment, $m(\vec{k})$ is composed of not only the local self-rotation [Eq. (C2)], but the center-of-mass motion of the wave packet as well [56], described as

$$m_{\text{orb}}(\vec{k}) = \frac{e}{2\hbar} \text{Im} \langle \partial_{\vec{k}} u(\vec{k}) | \times [H(\vec{k}) - \varepsilon(\vec{k})] | \partial_{\vec{k}} u(\vec{k}) \rangle$$
$$+ \frac{e}{\hbar} \text{Im} \langle \partial_{\vec{k}} u(\vec{k}) | \times [\varepsilon(\vec{k}) - E_F] | \partial_{\vec{k}} u(\vec{k}) \rangle. \quad \text{(C5)}$$

To compare the contributions from orbital and spin magnetization in our device, we calculate the components $\alpha_{xz}^{\text{orb}}$ and $\alpha_{xz}^s$. $\alpha_{xz}^{\text{orb}}$ ($\alpha_{xz}^s$) reflects the strength of induced $z$-direction orbital (spin) magnetization per unit electric field $E_x$. Figures 6(b) and 6(c) show the $\alpha_{xz}^{\text{orb}}$ and $\alpha_{xz}^s$ as functions of Fermi level for different temperatures, respectively. It can be seen that the $\alpha_{xz}^{\text{orb}}$ is about one to two orders of magnitude larger than $\alpha_{xz}^s$ when $\mu = 0.037$ eV (the Fermi level of the

TaIrTe$_4$ device), indicating the dominant contribution of orbital moment to the out-of-plane magnetization in our work.

### APPENDIX D: OBSERVED AHE WITH THE CURRENT ALONG THE CRYSTAL b AXIS

We perform measurements of the AHE with the bias current along the crystal $b$ axis of the TaIrTe$_4$ device, as shown in Fig. 7(a). Indeed, it is found that the amplitude of Hall resistance $R_H^\omega$ is much smaller when the $I^{\text{dc}}$ is injected along the $b$ axis, compared to the case of current along the $a$ axis [Fig. 7(b)]. With $I^{\text{dc}}$ along the $b$ axis where the current direction is perpendicular to Berry curvature dipole, the induced orbital magnetization is vanishingly small [22], which may explain the negligible Hall signals observed here.

### APPENDIX E: SECOND-ORDER NONLINEAR HALL EFFECT IN THE STUDIED DEVICE

The few-layer TaIrTe$_4$ is expected to possess a second-order nonlinear Hall effect (NLHE) due to its nonzero Berry curvature dipole along the crystal $a$ axis. The NLHE has been reported in a previous study [27] and also takes place in our TaIrTe$_4$ device. The measurement configuration of second-order NLHE is illustrated in Fig. 8(a). An alternating current $I^\omega$ is injected into the TaIrTe$_4$ flake from source (S) to drain (D) electrodes, and the longitudinal and transverse voltages are recorded at the fundamental and second-harmonic frequencies. The electrodes A and B for transverse signals are perpendicular to the current direction. Figure 8(b) shows the measured NLHE with the current applied along the crystal $a$ and $b$ axes, respectively. The $V_H^{2\omega}$ signal is almost zero when $I^\omega$ is injected along the $b$ axis due to the mirror symmetry $M_a$. The angular dependence of the NLHE is also measured in a rotating reference frame (S–A–D–B in a clockwise sequence) following the Hall measurement configuration. As shown in Fig. 8(c), the $V_H^{2\omega}$ scales linearly with $V_\parallel^2$ for different current directions. The slope $V_H^{2\omega}/V_\parallel^2$ versus the angle $\theta$ (current direction) is summarized in Fig. 8(d). The nonlinear Hall response reaches a maximum when the current is perpendicular to the mirror plane (that is, along the $a$ axis) but a minimum when the current is parallel to the mirror plane (along the $b$ axis). Due to the $Pm$ point group symmetry of $T_d$-TaIrTe$_4$ [66,68], the angular dependence of $V_H^{2\omega}/V_\parallel^2$ can be fitted through the formula $\frac{V_H^{2\omega}}{V_\parallel^2} = A\cos\theta \frac{d_{12}r^2\cos^2\theta + (d_{11}-2d_{26}r^2)\sin^2\theta}{(\sin^2\theta + r\cos^2\theta)^2}$ [19,27], where $d_{ij}$ are the elements of the second-order susceptibility tensor for the $Pm$ point group [78], $r$ is the resistance anisotropy defined as $r = R_a/R_b$, and $A$ is the overall amplitude. The good fitting [red curve in Fig. 8(d)] indicates that the $V_H^{2\omega}$ signals are fully consistent with the model of intrinsic crystal symmetry analysis of TaIrTe$_4$. On the other hand, this observation can rule out extrinsic asymmetries [20,35] as the origin of NLHE, such as the accidental diode at the electrode/sample interface, irregular sample geometry, and the thermoelectric effect due to Joule heating.

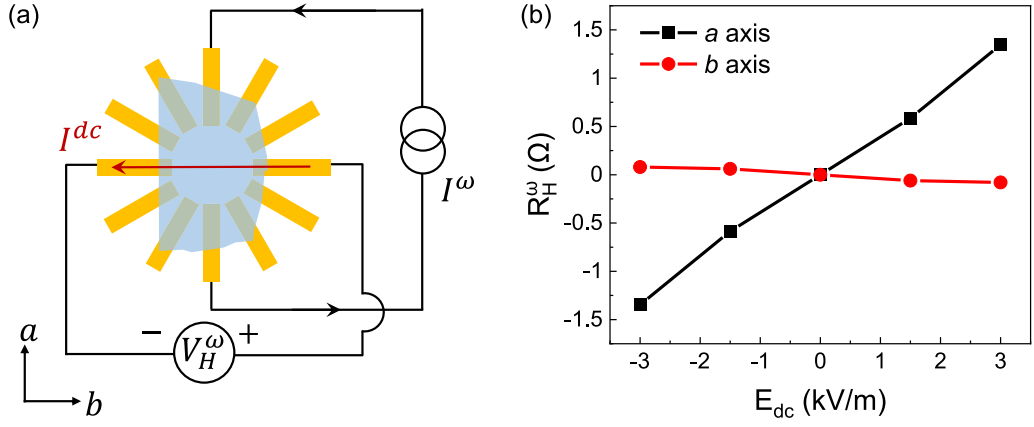

FIG. 7. Comparison of the AHE when the $I^{dc}$ is applied along different directions. (a) Schematic of measurement configuration for AHE with $I^{dc}$ along the $b$ axis. The direct current $I^{dc}$ is injected into the TaIrTe$_4$ flake along the $b$ axis, and the ac probe current $I^{\omega}$ is applied along the $a$ axis while recording the transverse voltage $V_H^{\omega}$. (b) The obtained Hall resistance $R_H^{\omega}$ with $I^{dc}$ applied along the $a$ and $b$ axes, respectively.

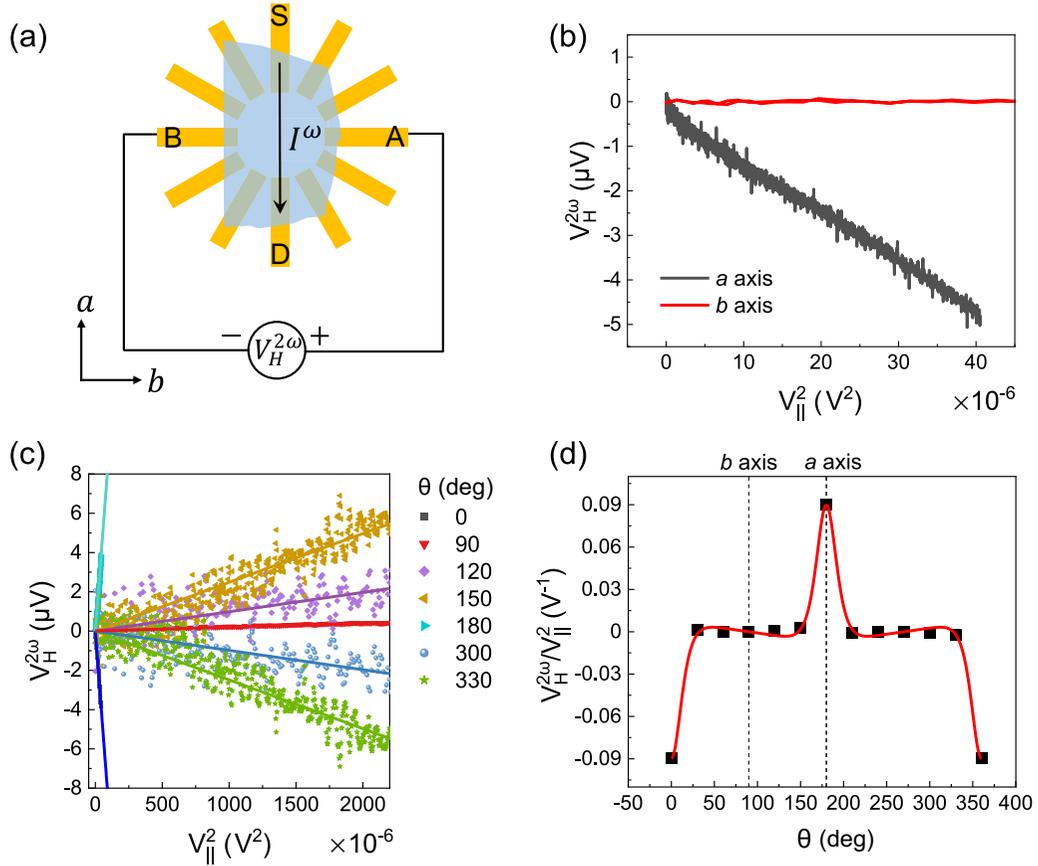

FIG. 8. Second-order nonlinear Hall effect and its angular dependence in the TaIrTe$_4$ device. (a) Schematic of measurement configuration for the nonlinear Hall effect. The ac current $I^{\omega}$ with driving frequency $\omega$ (17.777 Hz) is injected from the source (S) to drain (D) electrodes. The second-harmonic transverse voltage $V_H^{2\omega}$ is measured between the A and B electrodes. The measurement of first-harmonic longitudinal voltage $V_{\parallel}$ is not exhibited for simplicity. (b) The measured $V_H^{2\omega}$ vs the square of the first-harmonic longitudinal voltage $V_{\parallel}^2$ when the current is along the $a$ and $b$ axes, respectively. (c) $V_H^{2\omega}$ scales linearly with $V_{\parallel}^2$ for different current directions. The angle $\theta$ is defined as the relative orientation between $I^{\omega}$ and the $a$ axis. For example, the angle $\theta = 0°$ (90°) corresponds to the current along the $a$ axis ($b$ axis). (d) The nonlinear Hall effect [slope of the dependences in (c)] as a function of $\theta$. The black dots represent the experimental data, while the red curve is the fit to the experimental data with the formula described in the main text.

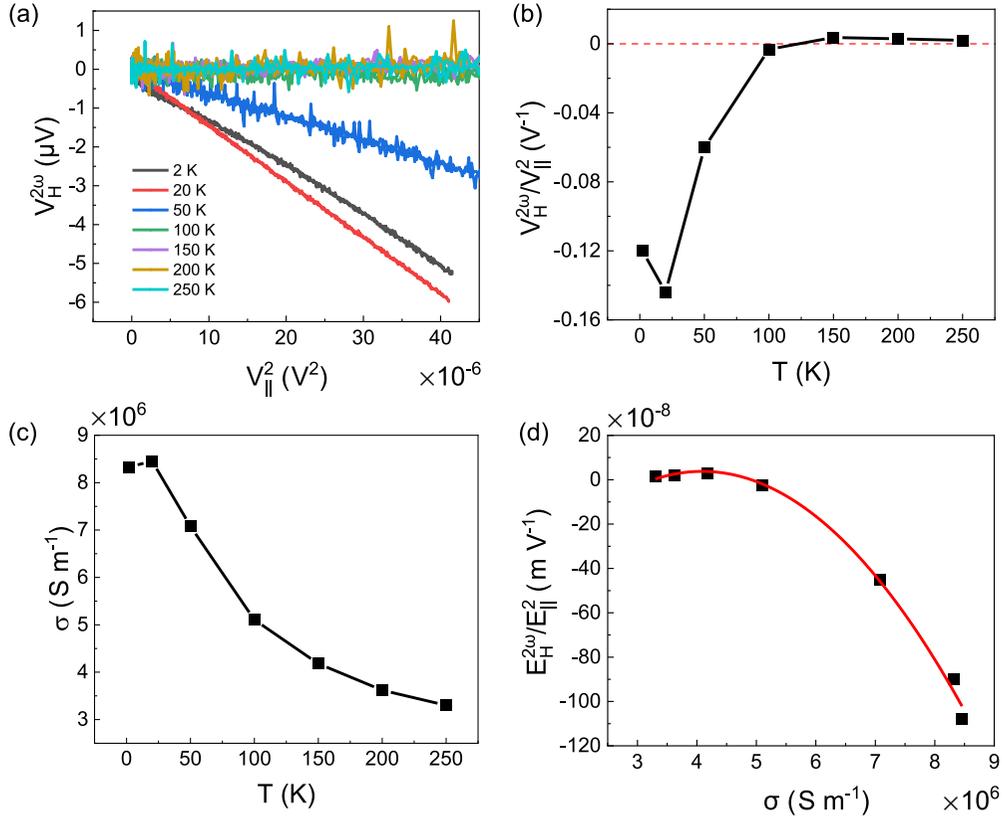

FIG. 9. Temperature dependence of nonlinear Hall effect in the TaIrTe$_4$ device with the current along the crystal $a$ axis. (a) $V_H^{2\omega}$ scales linearly with $V_\parallel^2$ at temperatures ranging from 2 to 250 K. (b) The slope of $V_H^{2\omega} - V_\parallel^2$ curves as a function of temperature. The slope reverses its sign when the temperature is raised above 100 K. (c) Temperature dependence of the longitudinal conductivity $\sigma$ of the TaIrTe$_4$ device. (d) The slope $E_H^{2\omega}/E_\parallel^2$ as a function of the conductivity $\sigma$. The black dots represent the experimental data, while the red curve is the parabolic fit to the experimental data.

We have also investigated the temperature dependence of NLHE in the TaIrTe$_4$ device, as shown in Fig. 9. The NLHE signal $V_H^{2\omega}$ depends linearly on the $V_\parallel^2$ at all temperatures [Fig. 9(a)]. The slope $V_H^{2\omega}/V_\parallel^2$ decreases with increasing the temperature from 20 to 100 K [Fig. 9(b)]. As the temperature is further increased, $V_H^{2\omega}/V_\parallel^2$ inverts its sign at ∼150 K and then approaches zero at 250 K. The conductivity exhibits a similar temperature dependence [Fig. 9(c)], where $\sigma$ gradually decreases with increasing $T$. To reveal the underlying mechanism of NLHE, we try to examine the scaling law of the NLHE in TaIrTe$_4$. Figure 9(d) shows the slope $E_H^{2\omega}/E_\parallel^2$ as a function of the conductivity $\sigma$. It can be seen that the NLHE is dominated by a parabolic behavior, indicating that both the intrinsic Berry curvature and extrinsic disorder scattering make contributions to the nonlinear Hall response [18,27,79]. However, notably, recent studies suggest that increasing temperature would induce a Fermi level shift in TaIrTe$_4$, leading to a sign inversion of the Berry curvature dipole [27]. The sign reversion of NLHE observed in Fig. 9(b) is likely to arise from the temperature-driven variation of the Fermi surface. Therefore, the validity of the scaling law analysis in Fig. 9(d) deserves more attention [18], which requires further investigation in the future.

## APPENDIX F: NONRECIPROCAL HALL EFFECT MEASURED VIA THE PURE DC METHOD

The AHE can be captured by a pure dc measurement method. The measurement configuration is given in Fig. 10(a), where $I^{dc}$ is applied along the $a$ axis, while the transverse voltage $V_H^{dc}$ is simultaneously recorded. Figure 10(b) provides the raw data of $V_H^{dc} - I^{dc}$ at different temperatures. At high temperatures, $V_H^{dc}$ is almost linear with $I^{dc}$, exhibiting a dominant antisymmetric behavior. However, as the temperature decreases, the $V_H^{dc} - I^{dc}$ curve begins to deviate from the antisymmetric linear behavior, which is especially prominent at 2 and 20 K. This is because the current-induced second-order transverse voltage is pronounced at low temperatures, leading to a parabolic contribution to the $V_H^{dc}$. By symmetrizing and antisymmetrizing the original $V_H^{dc} - I^{dc}$ data, we can obtain the symmetric component $V_H^s$ and antisymmetric component $V_H^{as}$, respectively. The antisymmetric component $V_H^{as}$ originates from the electrode misalignment, which scales linearly with the current $I^{dc}$ [Fig. 10(c)]. The symmetric component $V_H^s$, arising from the current-induced out-of-plane magnetization, exhibits a quadratic dependence on the current $I^{dc}$ [Fig. 4(d) in the main text]. The quadratic $V_H^s - I^{dc}$ relation suggests the emergence of colossal nonreciprocal Hall effect. Upon switching the direction of $I^{dc}$, the Hall resistance $R_H^s$ ($V_H^s/I^{dc}$) reverses its sign, thus leading to a divergent Hall nonreciprocity as defined by $\eta_H = \frac{R_H^s(I^{dc})-R_H^s(-I^{dc})}{R_H^s(I^{dc})+R_H^s(-I^{dc})} \to \infty$.

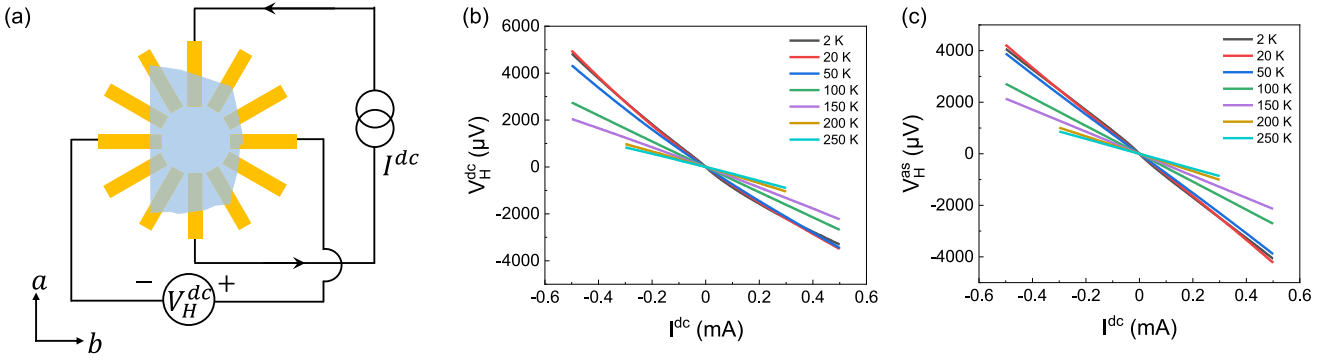

FIG. 10. The original $I^{dc}-V_H^{dc}$ data and its antisymmetric component measured at the TaIrTe$_4$ device. (a) Schematic of pure dc measurement for AHE. The dc current $I^{dc}$ is applied along the $a$ axis, while the transverse voltage $V_H^{dc}$ is simultaneously recorded along the $b$ axis. (b) The original $V_H^{dc} - I^{dc}$ data of the TaIrTe$_4$ device at different temperatures. (c) The antisymmetric component of $V_H^{dc}$, denoted as $V_H^{as}$, as a function of bias current $I^{dc}$. $V_H^{as}$ is obtained by antisymmetrizing the original $V_H^{dc} - I^{dc}$ data.

## APPENDIX G: COMPARISON OF AHE, NLHE, AND NRHE MEASURED IN THE DEVICE

As indicated by the discussion above and in the main text, the linear AHE, second-order NLHE, and NRHE in our TaIrTe$_4$ device share a similar origin, i.e., current-induced orbital magnetization through the nonzero Berry curvature dipole. However, the quantitative comparison between the corresponding experimental results requires extra attention. For example, an experimental value of the Berry curvature dipole can be obtained by analyzing the scaling behavior of the ratio $E_H^{2\omega}/(E^\omega)^2$ in NLHE measurements [18], but the same analysis on $E_H^\omega/(E^{dc}E^\omega)$ in AHE data may lead to incorrect results. Here, we briefly discuss the quantitative relation between these Hall effects, provide a valid approach to comparing these values, and reinforce our claims.

Within the regime of second-order electric response, all the above-mentioned Hall effects should follow the same expression,

$$V_H^{tot} = k(I^{tot})^2, \quad (G1)$$

where $I^{tot}$ includes any electrical current, either direct or alternating, along the $a$ axis of the device, i.e., $I^{tot} = I_1 + I_2 \sin \omega t$, with $I_1$ and $I_2$ representing the amplitudes of dc and ac excitation, respectively. $k$ is a general coefficient valid for all these Hall effects [12] and counts all mechanisms of second-order Hall response, namely, Berry curvature dipole, side-jump, and skew scattering. $V_H^{tot}$ is the overall second-order Hall response.

### 1. AHE with coexisting dc and ac excitation

In this case, both $I_1 = I^{dc}$ and $I_2 = I^\omega$ remain nonzero, and the AHE of our interest corresponds to the first-harmonic amplitude of Eq. (G1). Explicitly, we have $V_H^\omega \sin \omega t = 2kI^{dc}I^\omega \sin \omega t$. Thus, in AHE measurement, the generation of the Hall signal satisfies

$$\frac{V_H^\omega}{I^{dc}I^\omega} = 2k. \quad (G2)$$

### 2. Second-order NLHE with pure ac excitation

In this case, $I_1 = 0$ and $I_2 = I^\omega$, and the NLHE of our interest corresponds to the second-harmonic part of Eq. (G1). We have $V_H^{tot} = k(I^\omega)^2 \sin^2 \omega t = k(I^\omega)^2 \frac{1-\cos 2\omega t}{2}$, and the second-harmonic component becomes $V_H^{2\omega} = -\frac{1}{2}k(I^\omega)^2$. Despite the $\frac{\pi}{2}$ phase in the lock-in measurement setup, the generation of the NLHE signal then satisfies

$$\frac{V_H^{2\omega}}{(I^\omega)^2} = -\frac{k}{2}. \quad (G3)$$

### 3. NRHE with pure dc excitation

In this case, $I_1 = I^{dc}$ and $I_2 = 0$. Equation (G1) has no ac component and simply becomes $V_H^s = k(I^{dc})^2$, leading to the quadratic generation,

$$\frac{V_H^s}{(I^{dc})^2} = k. \quad (G4)$$

Equations (G2)–(G4) reveal the fact that when carrying out a quantitative comparison between experimental results of different Hall effects, an extra factor of 2 or $-1/2$ is required to arrive at a self-consistent outcome. This arises from either a current double counting (AHE) or the inherent structure of trigonometric functions (second-order NLHE). Specifically, the above ratios should satisfy the relation $V_H^\omega/(I^{dc}I^\omega)$ : $V_H^{2\omega}/(I^\omega)^2$ : $V_H^s/(I^{dc})^2 = 4 : -1 : 2$. We convert these quantities to the more sample-independent values $E_H^\omega/(E^{dc}E^\omega)$, $E_H^{2\omega}/(E^\omega)^2$ and $E_H^s/(E^{dc})^2$ by multiplying a $\frac{L}{R_a^2}$ factor, where $L$ and $R_a$ are the channel length and resistance along the $a$ axis of the sample, respectively. In Fig. 11, we quantitatively compare these corrected values by plotting them in pairs with respect to temperature. This good consistency provides further evidence of the Berry curvature dipole origin of the AHE in our work.

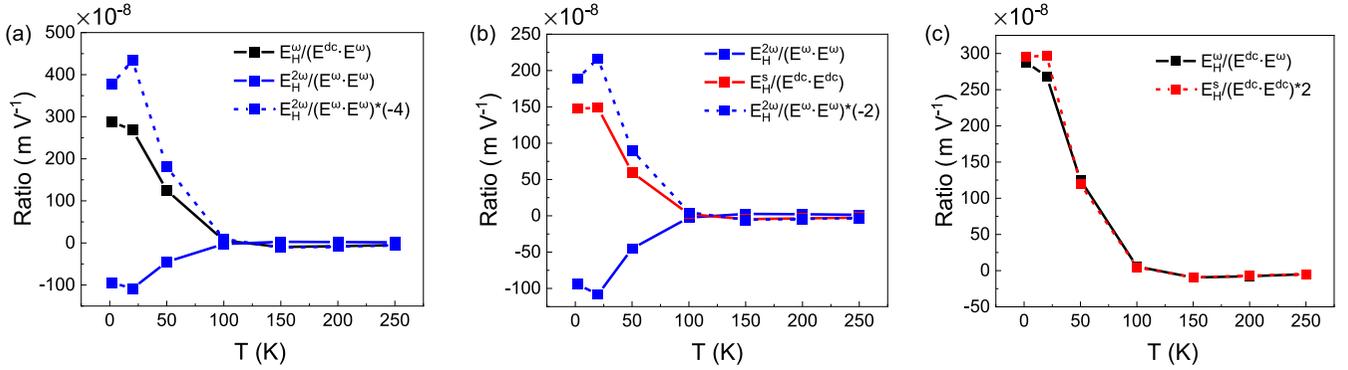

FIG. 11. Comparison of the AHE, second-order NLHE, and NRHE observed in the TaIrTe$_4$ device. The generation ratio of the Hall signal for the AHE (black), second-order NLHE (blue), and NRHE (red) is denoted by $E_H^\omega/(E^{dc}E^\omega)$, $E_H^{2\omega}/(E^\omega E^\omega)$, and $E_H^s/(E^{dc}E^{dc})$, respectively. It is found that the measured $E_H^\omega/(E^{dc}E^\omega)$ is almost larger than that of $E_H^{2\omega}/(E^\omega E^\omega)$ by a factor of $-4$, while the values of $E_H^\omega/(E^{dc}E^\omega)$ and $E_H^s/(E^{dc}E^{dc})$ have a difference by a factor of $\sim 2$.

## APPENDIX H: DOMINANT ROLE OF ORBITAL MAGNETIZATION IN THE OBSERVED AHE

Both the orbital and spin magnetization could possibly lead to the observation of AHE in principle. We think the current-induced AHE in our work (Figs. 3 and 4) is mainly originated from the orbital magnetic moment (instead of spin) according to the following aspects.

### 1. Current and angle dependences of AHE

The current and angle dependences of dc-induced Hall effect are both consistent with the scenario of orbital magnetic moment. The Hall resistance exhibits a linearity with the dc current bias applied along the $a$ axis [Fig. 3(c)], in accordance with the linear relationship between orbital magnetization and dc current. Besides, the Hall resistance is found vanishingly small when the dc current is applied along the $b$ axis (Fig. 7), which is expected due to the negligible orbital magnetization as the current is perpendicular to Berry curvature dipole in TaIrTe$_4$.

### 2. Good consistency with the second-order NLHE

The observed linear Hall effect induced by a dc current is fully consistent with the results of second-order NLHE induced by an ac current in our TaIrTe$_4$ device. For example, the temperature-induced sign reversal is observed in both measurement methods [Figs. 4(b) and 4(c)], and the generation ratio of the Hall signal also agrees with each other [Fig. 11(a)]. This good consistency indicates that the observed linear AHE has the same origin as that of second-order NLHE, which has been extensively investigated [19,20,27] and is believed to stem from the current-induced orbital magnetization via the Berry curvature dipole.

### 3. Theoretical comparison of orbital and spin magnetization

It is noteworthy that, owing to the presence of spin-orbit coupling in TaIrTe$_4$, the orbital magnetization gives rise to a spin polarization, leading to a corresponding spin anomalous Hall effect. While it can be challenging to disentangle the contributions of orbit and spin in systems with strong spin-orbit coupling, the fundamental essence of this observation is rooted in the orbital effect. The spin magnetization $M_s^z$, a product of spin-orbit coupling, is notably smaller than the orbital magnetization $M_{orb}^z$, as indicated by calculations [Fig. 2(d)]. In the AHE scenario, the Hall resistance $R_H$ is generally proportional to the magnetization $M^z$ described by the formula $R_H = \gamma M^z$ with $\gamma$ a sample-dependent parameter. Therefore, in our device, the observed Hall effect (e.g., $R_H$) should be dominantly contributed by the orbital magnetic moment instead of the spin magnetic moment.